\documentclass[conference]{IEEEtran}
\IEEEoverridecommandlockouts
\usepackage{cite}
\usepackage{amsmath,amssymb,amsfonts,amsthm}
\usepackage{hyperref}
\usepackage{algorithmic}
\usepackage{graphicx}
\usepackage{textcomp}
\usepackage{xcolor}
\usepackage{caption}

\def\BibTeX{{\rm B\kern-.05em{\sc i\kern-.025em b}\kern-.08em
    T\kern-.1667em\lower.7ex\hbox{E}\kern-.125emX}}
\definecolor{darkgreen}{rgb}{0,0.39,0}

\DeclareMathOperator{\Tr}{Tr}

\begin{document}

\bstctlcite{IEEEexample:BSTcontrol}

\title{Physics-informed and Unsupervised Riemannian Domain Adaptation for Machine Learning on Heterogeneous EEG Datasets\\
\thanks{This work was supported by the grants ANR-20-CHIA-0016, ANR-20-IADJ-0002, and ANR-20-THIA-0013 from Agence nationale de la recherche (ANR).}
}

\author{\IEEEauthorblockN{Apolline Mellot\IEEEauthorrefmark{2}\IEEEauthorrefmark{1}, Antoine Collas\IEEEauthorrefmark{2}, Sylvain Chevallier\IEEEauthorrefmark{3}, Denis Engemann\IEEEauthorrefmark{4}, Alexandre Gramfort\IEEEauthorrefmark{2}}
\IEEEauthorblockA{\IEEEauthorrefmark{2}University Paris-Saclay, Inria, CEA, Palaiseau, France. \IEEEauthorrefmark{3}TAU Inria, LISN-CNRS, University Paris-Saclay, France. \\
\IEEEauthorrefmark{4}Roche Pharma Research and Early Development, Neuroscience and Rare Diseases, \\ Roche Innovation Center Basel, F. Hoffmann–La Roche Ltd. \IEEEauthorrefmark{1}Email: apolline.mellot@inria.fr
}
}

\maketitle

\begin{abstract}

Combining electroencephalogram (EEG) datasets for supervised machine learning (ML) is challenging due to session, subject, and device variability. ML algorithms typically require identical features at train and test time, complicating analysis due to varying sensor numbers and positions across datasets. Simple channel selection discards valuable data, leading to poorer performance, especially with datasets sharing few channels. To address this, we propose an unsupervised approach leveraging EEG signal physics. We map EEG channels to fixed positions using field interpolation, facilitating source-free domain adaptation. Leveraging Riemannian geometry classification pipelines and transfer learning steps, our method demonstrates robust performance in brain-computer interface (BCI) tasks and potential biomarker applications. Comparative analysis against a statistical-based approach known as Dimensionality Transcending, a signal-based imputation called ComImp, source-dependent methods, as well as common channel selection and spherical spline interpolation, was conducted with leave-one-dataset-out validation on six public BCI datasets for a right-hand/left-hand classification task. Numerical experiments show that in the presence of few shared channels in train and test, the field interpolation consistently outperforms other methods, demonstrating enhanced classification performance across all datasets. When more channels are shared, field interpolation was found to be competitive with other methods and faster to compute than source-dependent methods.

\end{abstract}
\section{Introduction}
Electroencephalography (EEG) has emerged as a valuable tool for brain health monitoring and neuroscience studies due to its high temporal resolution and non-invasive nature \cite{niedermeyer2005electroencephalography}. Supervised machine learning algorithms have driven significant advances in EEG signal analysis, enabling applications such as sleep staging \cite{perslev2021u}, pathology diagnostics \cite{gemein2020machine}, and brain-computer interfaces (BCIs) \cite{clerc2016brain}.

While machine learning is a promising tool for large-scale EEG analysis, its effectiveness can be compromised by dataset shifts. This problem occurs when a model is applied to unseen data (target) with a different distribution than the data on which it was trained (source) \cite{dockes2021preventing}. In EEG data, this variability arises from inter-subject differences, inter-session variability, and differences in recording equipment, e.g., amplifiers and electrode nets. To address this challenge, domain adaptation aims to bridge the gap between source and target datasets, thereby reducing the impact of dataset shift \cite{bleuze2021transfer, altindis2023transfer}. In addition to dataset shift, EEG data often have varying numbers and positions of electrodes in different recording setups, resulting in heterogeneous data. 

Various methods have been proposed to address the challenge of dimension mismatch in domain adaptation for EEG data. Methods based on projecting data into lower-dimensional spaces \cite{tveitstol17introducing} or selecting common channels \cite{wei20222021, mellot} risk losing valuable information specific to EEG signals. End-to-end deep learning approaches propose to leverage spatial attention mechanisms by using the electrodes' coordinates \cite{truong2023deep} but are computationally expensive. While covariance-based imputation approaches like Dimensionality Transcending \cite{rodrigues2020dimensionality} achieve state-of-the-art performance in BCI applications, they rely on supervised learning, limiting their application to scenarios with labeled target data.

In this work, we propose an unsupervised source free pipeline to combine heterogeneous datasets: a model is trained on several labeled datasets (sources) and then applied on an unseen and unlabeled dataset (target). We propose to map EEG channels to a template of fixed positions by leveraging the underlying physics of EEG data through field interpolation. Then, we leverage the Riemannian re-centering operator to align the statistical distributions of data from different source domains, mitigating dataset shift. We evaluated our method on six BCI datasets using leave-one-dataset-out procedures. 

This paper is organized as follows: Section II presents the Riemannian classification pipeline, Section III introduces the methods explored in this work to deal with heterogeneous EEG data, and Section IV describes the experimental evaluation of these methods and reports on the advantages of field interpolation compared to other methods.
\begin{figure}
    \centering
    \begin{tabular}{ l c c c c }
        \hline
        \textbf{Datasets} & \textbf{Subj.} & \textbf{Chan.} & \textbf{Sess.} & \textbf{Runs} \\ 
        \hline 
        BNCI2014\_001 (B1)& 12 & 22 & 2 & 6 \\
        BNCI2014\_004 (B4)& 9 & 3 & 5 & 1 \\ 
        PhysionetMI (P)& 109 & 64 & 1 & 1 \\ 
        Shin2017A (S)& 29 & 30 & 3 & 1 \\ 
        Weibo2014 (W)& 10 & 60 & 1 & 1 \\ 
        Zhou2016 (Z)& 4 & 14 & 3 & 2 \\ 
        \hline 
    \end{tabular}
    \centering
    \includegraphics[width=\linewidth]{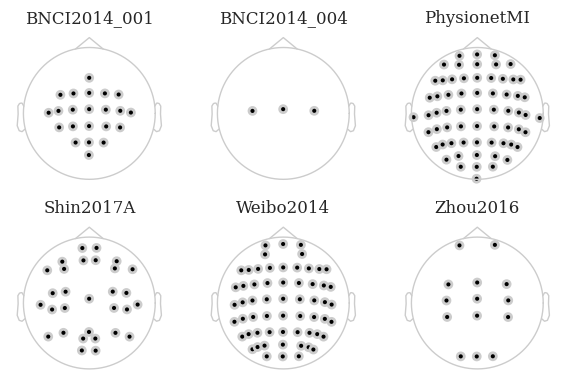}
    \caption{Top: table of diverse characteristics of the datasets. Bottom: 2D projection of sensor positions on the scalp.}
    \label{fig:positions}
\end{figure}
\section{EEG covariance and Riemannian geometry}
\subsection{Covariance matrix representation}
\label{subec:covs}
EEG signals $\textbf{X} \in \mathbb{R}^{P \times T}$ are multivariate time series recorded with $P$ sensors that capture the electrical activity of the brain at $T$ time steps. We represent the EEG signal by its empirical covariance matrix:
\begin{equation}
    \textbf{C} = \frac{\textbf{X}\textbf{X}^{\top}}{T} \in \mathbb{R}^{P \times P} \enspace .
\end{equation}
As we deal with interpolated signals filled with linear combinations of the sensors' signals, we regularised the empirical covariance matrices with the Ledoit-Wolf shrinkage to avoid rank deficiency: $\hat{\textbf{C}} = (1-s) \textbf{C} + s \mu \mathbf{I}_P$
with $\mu = \Tr(\textbf{C}) / P$ and $s$ the shrinkage coefficient \cite{ledoit2004well}. The regularised covariance matrix belongs to the set $\mathcal{S}^{++}_P$ of Symmetric Positive Definite (SPD) matrices. 
\subsection{Riemannian geometry}
SPD matrices lie in a manifold that can be equipped with the affine-invariant Riemannian distance \cite{forstner2003metric}:
\begin{align} \label{eq:riemann_dist}
    \delta_R(\textbf{C}_1, \textbf{C}_2) = \left( \sum_{k=1}^{P} \text{log}^2 \lambda_{k}\right)^\frac{1}{2}\enspace ,
\end{align}
where $\lambda_{k}$ are the eigenvalues of $\textbf{C}_1^{-1} \textbf{C}_2$.
With this distance, we define the geometric Riemannian mean of a set of $N$ SPD matrices by minimizing:
\begin{align} \label{eq:riemannian_mean}
    \bar{\textbf{C}} = \operatorname*{argmin}_{\textbf{C} \in \mathcal{S}_P^{++}} \sum_{i=1}^{N} \delta_{R}^{2}(\textbf{C}, \textbf{C}_i) \enspace .
\end{align}
Endowed with a Riemannian metric, a Riemannian manifold $\mathcal{M}$ is a differentiable manifold to which we can attach a Euclidean vector space at any point of the manifold, called tangent space. Each element of the manifold can be projected on the tangent space by a logarithmic mapping. The tangent vector $\textbf{z}$ is the vectorized projection of a SPD matrix $\textbf{C}$ on the tangent space at point $\bar{\textbf{C}}$:
\begin{align}
    \textbf{z} = \textrm{Upper}(\textrm{log}(\bar{\textbf{C}}^{-\frac{1}{2}} \textbf{C} \bar{\textbf{C}}^{-\frac{1}{2}})) \in \mathbb{R}^{P(P+1)/2} \enspace ,
    \label{eq:vector}
\end{align}
where $\textrm{Upper}()$ is the function returning a vector containing the concatenation of the upper triangle values of a matrix with weights of 1 for diagonal elements and $\sqrt{2}$ for other elements, and $\textrm{log}()$ the matrix logarithm with the log applied to the eigenvalues in the eigenvalue decomposition. This tangent space mapping transforms covariance matrices into Euclidean vectors that can be used as input for classical linear machine learning models \cite{barachant}.
\subsection{Transfer learning: re-center to Identity}
\label{subsec:recenter}
In covariance-based BCI classification, the preferred transfer learning technique to reduce the shifts induced by subject and dataset variability is re-centering the covariance distributions to a common reference point on the Riemannian manifold \cite{zanini2018}. Here, each covariance distribution, or domain, is re-centered to the Identity by whitening them with their respective geometric mean $\bar{\textbf{C}}$:
\begin{align}
    \textbf{C}_i^{(\text{rct})} = \bar{\textbf{C}}^{-\frac{1}{2}} \textbf{C}_i \bar{\textbf{C}}^{-\frac{1}{2}} \enspace .
\end{align}
One whitening is applied per domain separately. When applied to EEG data, domains can be defined as the datasets, the subjects, or even the sessions. In our experiments, we consider each subject as one domain.
\section{Matching EEG data dimensions}
\label{sec:match_dim}
This work focuses on combining several heterogeneous datasets with varying numbers of channels due to different recording devices. \autoref{fig:positions} illustrates the variability in positions of the sensors on the scalp across the six datasets considered in the experimental evaluation. We present existing strategies to deal with heterogeneous dimensionalities and propose a new source free approach with channel interpolation.
\subsection{Common channel selection}
Selecting EEG channels that are consistently present across all subjects and datasets is the first strategy that comes to mind, particularly effective when there is a sufficient number of shared channels to perform the desired analytical tasks \cite{wei20222021, mellot}. However, as the number of subjects and datasets increases, the variability in sensor positions due to the use of different recording devices becomes more pronounced. This can lead to a situation where there are too few common channels for the task, as was observed in our experimental evaluation, where only one common channel in the middle of the head (Cz) was insufficient for right/left-hand classification.
\subsection{Geometry-based imputation: Dimensionality Transcending (DT)}
Rodrigues et al. \cite{rodrigues2020dimensionality} proposed to transform the data points into an expanded common space with an isometric transformation, which preserves the distance of the original data points in the new expanded space. We consider $M$ datasets of different dimensionality (or different numbers of sensors) $P_j$ with $j=1, \dots, M$. The expanded space dimension is $P_{\textrm{exp}}= \bigcup P_j$. The resulting  expanded version $\textbf{C}_i^\uparrow$ of a matrix $\textbf{C}_i$ from the $j$-th dataset is:
\begin{align}
    \textbf{C}_i^\uparrow = 
    \begin{bmatrix}
        \textbf{C}_i & \textbf{0}_{P_j\times(P_{\textrm{exp}}-P_j)} \\
        \textbf{0}_{(P_{\textrm{exp}}-P_j)\times P_j} & \mathbf{I}_{(P_{\textrm{exp}}-P_j)}
    \end{bmatrix} \in \mathbb{R}^{P_{\textrm{exp}} \times P_{\textrm{exp}}}
\end{align}
This transformation is applied to all datasets to get covariances of size $P_{\textrm{exp}} \times P_{\textrm{exp}}$. In addition, a permutation of the expanded matrices rows and columns is performed to make the same channels correspond across datasets. This expansion of covariances was originally designed to be followed by a supervised transfer learning method called Riemannian Procrustes Analysis (RPA) \cite{rodrigues2018riemannian}. However, since we operate in an unsupervised setting, we only apply the re-centering step introduced in \autoref{subsec:recenter} in our framework.
\subsection{Signal-based imputation}
The ``ComImp" method \cite{nguyen2022combining}, offers a direct machine learning approach to handle missing data in EEG recordings. First, the time series data are expanded to match the union of the set of channels, hence generating missing values in the data. In our case the EEG signal is expanded to be of size $P_{\text{exp}} \times T$. Then, statistical imputation techniques can be employed. Such approaches require fitting the imputation method on training data that include at least one non-missing value per channel. Here, we considered a multivariate imputer that models each feature with missing values as a function of the other features with a regression. We used the IterativeImputer class of the Scikit-Learn software \cite{scikit-learn} with a ridge estimator.
\subsection{Physics-informed approach: Interpolation}
\begin{figure}
    \centerline{\includegraphics[width=\linewidth]{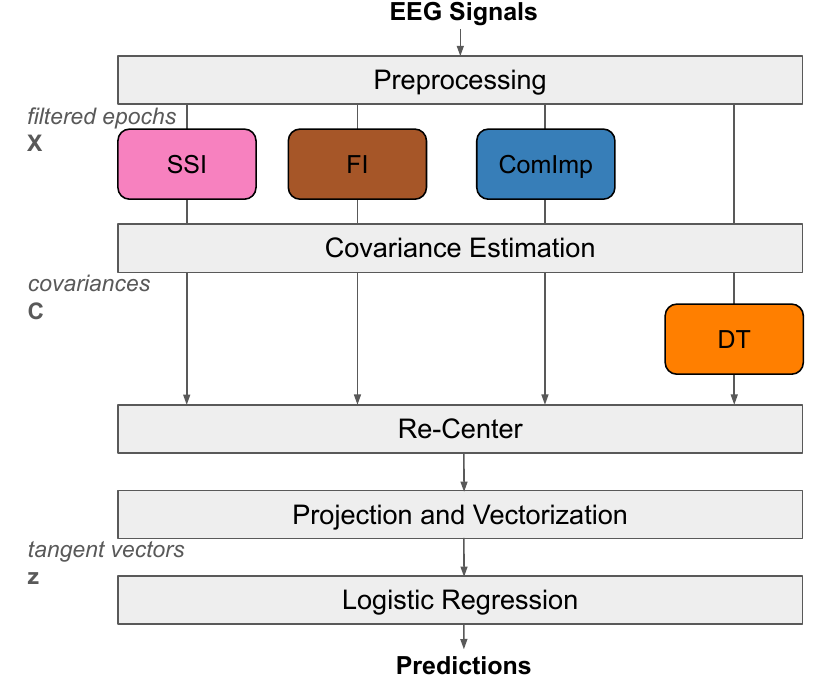}}
    \caption{Processing pipeline of EEG data. Depending on the method, dimensions are matched either when data are represented as epochs for interpolations or as covariances for DT. When there is no alignment performed, the Re-Center step is removed.}
    \label{fig:pipeline}
\end{figure}
Interpolation is a technique used in the context of EEG data processing to reconstruct the signals of malfunctioning or too noisy channels, usually referred to as `bad' channels. It uses the signals from the functional channels around the bad ones.
In this work, we used interpolation to map the different channels of EEG datasets: the EEG signal was reconstructed on fixed final positions based on the existing signal from all sensors of the datasets. Interpolation involves constructing a linear operator $\textbf{A} \in \mathbb{R}^{P \times P_j}$ that maps the $P_j$ existing EEG channels to the $P$ positions of a fixed template: $\hat{\textbf{X}} = \textbf{AX}$. $\textbf{X} \in \mathbb{R}^{P_j \times T}$ are the recorded EEG signals and $\hat{\textbf{X}} \in \mathbb{R}^{P \times T}$ are the reconstructed signals. This operator can be estimated to reconstruct the EEG signal at any desired position, even if there is no corresponding sensor at that location. 
Depending on the EEG montage, $P$ can either be smaller or larger than $P_j$.
We present two interpolation techniques used with EEG data: the spherical spline interpolation (SSI) and the field interpolation (FI).

\paragraph*{SSI}
The idea behind SSI is to model the data using smooth functions that are defined on the surface of a sphere. The functions here are spherical splines~\cite{PERRIN1989184}. In practice, the existing sensors' locations and desired final positions are first projected onto a unit sphere. Then, the linear mapping matrix is computed and finally used to interpolate the signal at the desired position based on the existing signal. 
\begin{figure*}[t]
    \centerline{\includegraphics[width=\textwidth]{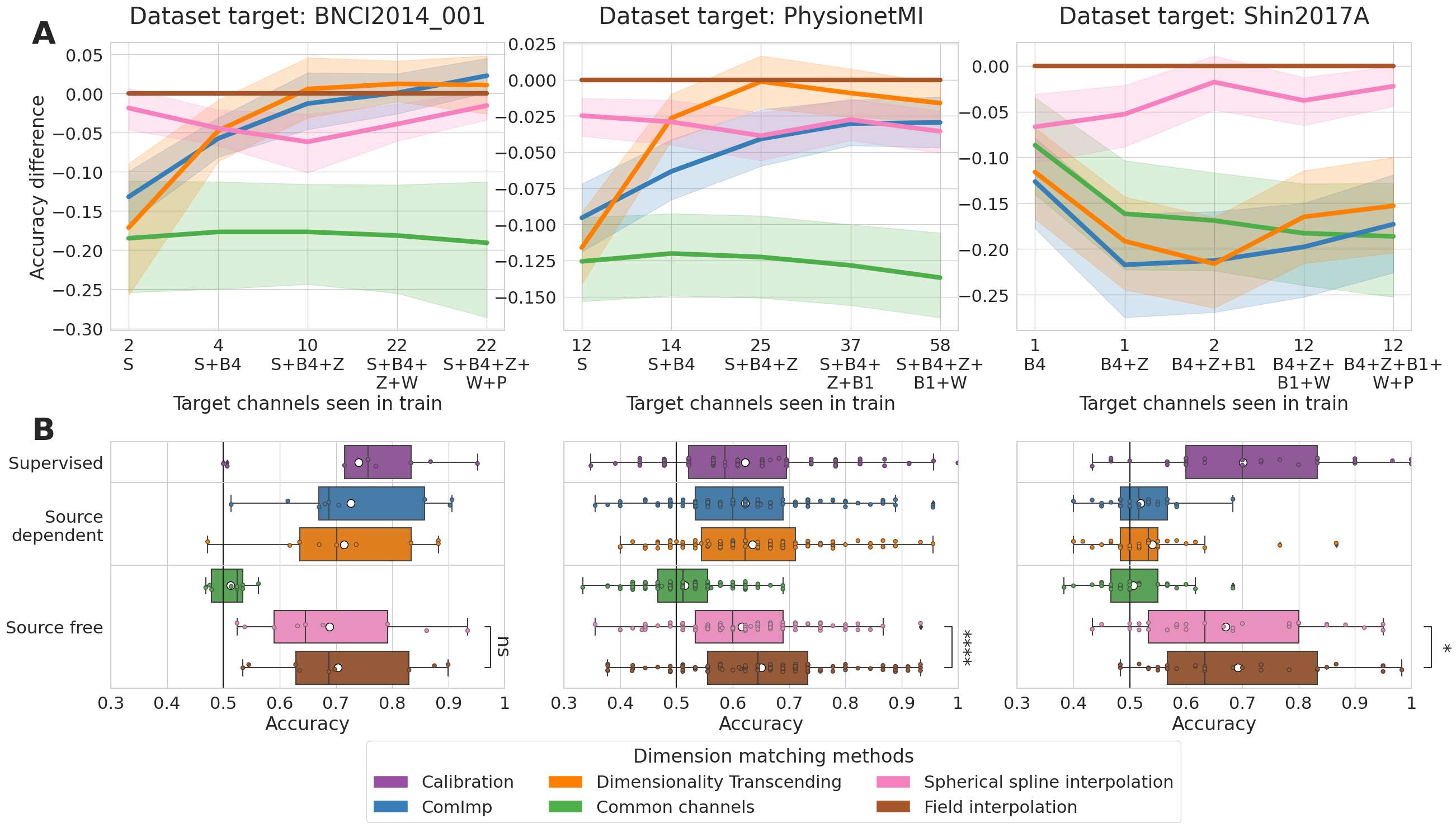}}
    \caption{Accuracy for different dimensionality matching methods on three left-out datasets. One column (across panels \textbf{A} \& \textbf{B}) corresponds to one target dataset. (\textbf{A}) Comparative learning curves for an increasing number of target channels seen during training with FI performance as reference. The increasing number of seen target channels is obtained by gradually including datasets in the train set, which is specified in the x-axis. The error bar represents the 95\% confidence interval over the target subjects' performance. (\textbf{B}) Boxplot of accuracies when the classifier is trained on the five other datasets. One point corresponds to one subject of the target dataset. A black line represents the median of the box and the mean by a white circle. The black lines indicate the chance level. The stars represent the results of a Wilcoxon test (ns: $p > 5\mathrm{e}{-2}$, *: $1\mathrm{e}{-2} < p \leq 5\mathrm{e}{-2}$, ****: $p \leq 1\mathrm{e}{-4}$).}
    \label{fig:results}
\end{figure*}
\paragraph*{FI}
The field mapping employed in the FI uses a canonical distribution of signal generators in the brain and then estimates a forward model based on Maxwell's equations \cite{GramfortEtAl2013a}. The forward model relates the distribution of estimated brain activity to sensor data at any electrode location. To obtain electric potential estimates at a missing electrode location, the available data are mapped to brain space using a Tikhonov regularization referred to as Minimum Norm Estimate (MNE) \cite{hamalainen1994interpreting}. Then the forward model can be applied to the estimated EEG generators to obtain potential values for any electrode location. The MNE-python software provides an implementation of the FI \cite{GramfortEtAl2013a}.
\section{Experimental evaluation}
\subsection{Data description}
The methods described in \autoref{sec:match_dim} were evaluated on six publicly available BCI datasets from the MOABB repository \cite{Aristimunha_Mother_of_all_2023}. The number of subjects, channels, sessions, runs, and trials varies across the datasets. \autoref{fig:positions} summarizes the key details of the datasets. All datasets were recorded with the left mastoid as reference, except for \emph{Weibo2014} where the reference was taken at the nose. All datasets consist of EEG data recorded while participants performed motor imagery tasks. The subjects were instructed to imagine moving either their right or left hand without actually moving it in response to a visual cue. The classification problem is thus a binary classification task.
\subsection{Preprocessing and classification pipeline}
All data were band-pass filtered between 8 and 32\,Hz with an Infinite Impulse Response (IIR) forward-backward filter and resampled at 128\,Hz. The signals were segmented in epochs at each trial and of duration corresponding to the trial length with no overlap. Subsequently, spatial covariance matrices were computed from the filtered epochs as presented in \autoref{subec:covs}. \autoref{fig:pipeline} illustrates the classification pipeline depending on the dimension-matching method. Subjects are used as domains for re-centering. The geometric mean of the subjects used in the train set is computed by considering all their data points. For the test subject, the geometric mean is computed on the data from its first session, or first run if there is only one session, or the first half of the data if there is only one session and one run.

The six datasets we used have only one channel in common: the Cz channel, located at the top of the head. The common channel selection thus resulted in keeping the signal from the Cz channel, and the associated covariance matrices came down to one value, the variance of this channel. As the union of the channels across all datasets represents $P_{\text{exp}}=84$ channels, the time series were expanded to 84 channels after ComImp, and the expanded covariance matrices after DT were of size $84 \times 84$. 
For both interpolation methods, we set the final positions to which all epochs across all datasets were interpolated to the following $P=17$ channels: [Fp1, Fp2, F7, F3, Fz, F4, F8, C3, Cz, C4, P3, Pz, P4, T3, T4, T5, T6]. 
Even though we chose $P < P_{\text{exp}}$, all channels of each dataset were used to reconstruct the interpolated signals. We determined this number of final positions to ensure that the task-related data was present in the reconstructed signals.
The regularization term of the SSI was set to $1\mathrm{e}{-7}$ and the one of the FI to $1\mathrm{e}{-3}$. For all methods, we used as classifier a logistic regression from Scikit-Learn with a L2 regularization set to $C=1$.
In addition, we employed a subject-specific calibration procedure to estimate the upper bound of achievable performance. Here we use a fraction of the target subject's data as part of the training set. This was achieved by splitting the target subject's data into two halves, utilizing the first half for training and the second for evaluation. Consequently, this yielded a single calibration accuracy value for each subject.
\subsection{Leave-one-dataset-out validation}
We evaluated the dimension-matching methods using a leave-one-data-out scheme. We first performed this evaluation with an increasing number of target channels seen in the training set by increasing the number of training datasets. In a second step, five of the six datasets were combined to form the training set, including all of their subjects. Each subject of the left-out dataset was used as a test set, resulting in one performance value per subject and method. This procedure was repeated six times so that each dataset was left out once. To statistically evaluate the difference in performance between SSI and FI, we performed a Wilcoxon signed-rank test on the classification accuracies obtained from ingle-trial predictions per method.

\subsection{Results}
Results for the three datasets with the most subjects are shown in \autoref{fig:results}. The first row (\textbf{A}) illustrates the classification accuracy difference of each subject of the target dataset as a function of the number of target channels present in the training datasets. The difference was made by subtracting the FI accuracy to other methods, subject per subject. For BNCI2014\_001 and PhysionetMI, ComImp and DT reach lower accuracies than FI with few target channels seen during training. When more target channels are seen in train, source free methods perform similarly to source dependent methods, with accuracies comparable to those of the calibration (purple box on row (\textbf{B})). For Shin2017A, FI consistently outperforms all other methods, especially ComImp, DT and common channel selection. The reason is that Shin2017A shares very few channels with the other datasets, and these channels are not located near the motor cortex. The boxplots of the second row (\textbf{B}) correspond to the last point of the learning curves, for which five of the datasets were used as training set. The methods are sorted according to whether they require access to target labels (supervised), source data (source dependent), or nothing (source free). For BNCI2014\_001 and PhysionetMI, interpolation methods performed similarly to source dependent methods and calibration. For Shin2017A, the observations are the same as on the learning curve. Out of all six datasets, the FI led to accuracies significantly higher than the SSI for four datasets with p-values $p \leq 5\mathrm{e}{-2}$. In addition, interpolation methods were faster to compute than ComImp and DT due to the smaller size of the dimension matched data (17 channels for interpolation compared to 84 for ComImp and DT). Overall, it is important to note that FI achieved similar or better accuracies than ComImp and DT with less information.
\section{Conclusion}
In this work, we propose a physics-informed, unsupervised, and source free domain adaptation approach for EEG analysis, specifically addressing challenges arising from varying electrode configurations. Leveraging the underlying physics of EEG signals with field interpolation led to performance similar to supervised and source dependent approaches when a large variety of data is available. Our approach showed to be better than other methods when the training data included limited target channels. In addition, interpolation is applied to the raw data before any feature extraction. It can therefore be flexibly used with different EEG pipelines and is not confined to the covariance framework.
\bibliographystyle{IEEEtran}
\bibliography{biblio}
\end{document}